\documentclass[useAMS,usenatbib,usegraphicx]{mn2e}

\title[Assembly of massive galaxies since $z$ = 1]
{Witnessing the active assembly phase of massive galaxies since $z$ = 1}
\author[Y. Matsuoka and K. Kawara]{Y. Matsuoka$^{1,2}$\thanks{Research Fellow of the Japan Society for the Promotion 
of Science}\thanks{E-mail: matsuoka@a.phys.nagoya-u.ac.jp} and K. Kawara$^{2}$\\
$^{1}$Graduate School of Science, Nagoya University, Furo-cho, Chikusa-ku, Nagoya 464-8602, Japan\\
$^{2}$Institute of Astronomy, The University of Tokyo, Osawa 2-21-1, Mitaka, Tokyo 181-0015, Japan}
\begin{document}

\date{Accepted 2010 February 1. Received 2010 January 21; in original form 2009 April 27}

\pagerange{\pageref{firstpage}--\pageref{lastpage}} \pubyear{2010}

\maketitle

\label{firstpage}

\begin{abstract}
We present an analysis of $\sim$60 000 massive (stellar mass $M_{\star} > 10^{11} M_{\odot}$) galaxies out to $z = 1$ 
drawn from 55.2 deg$^2$ of the United Kingdom Infrared Telescope (UKIRT) Infrared Deep
Sky Survey (UKIDSS) and the Sloan Digital Sky Survey (SDSS) II Supernova Survey. This
is by far the largest survey of massive galaxies with robust mass estimates, based on infrared
($K$-band) photometry, reaching to the Universe at about half its present age. We find that
the most massive ($M_{\star} > 10^{11.5} M_{\odot}$) galaxies have experienced rapid growth in number since
$z = 1$, while the number densities of the less massive systems show rather mild evolution. Such
a hierarchical trend of evolution is consistent with the predictions of the current semi-analytic
galaxy formation model based on $\Lambda$CDM theory. While the majority of massive galaxies are
red-sequence populations, we find that a considerable fraction of galaxies are blue star-forming
galaxies. The blue fraction is smaller in more massive systems and decreases toward the local
Universe, leaving the red, most massive galaxies at low redshifts, which would support the
idea of active 'bottom-up' formation of these populations during $0 < z < 1$.
\end{abstract}

\begin{keywords}
  galaxies: elliptical and lenticular, cD -- 
  galaxies: evolution --
  galaxies: formation --
  galaxies: luminosity function, mass function --
  galaxies: stellar content --
  cosmology: observations.
\end{keywords}

\section{Introduction}

Understanding the origin and evolution of galaxies, in particular
the most massive, is one of the major challenges in modern astrophysics.
Many massive galaxies today are giant early-type systems;
hence the formation of spheroids should proceed to a certain extent
in locked step with the mass assembly. The compelling theory of
hierarchical galaxy formation predicts that galaxies are assembled
through successive mergers of smaller systems in overdensities,
or haloes, of hypothetical cold dark matter (CDM) \citep{white78}.
Massive galaxies therefore emerge in the last phase of the formation
history. Alternatively, massive galaxies could form through
the rapid collapse of gas followed by a single prominent starburst at
high redshifts \citep{eggen62, larson75}.
This 'monolithic' scenario is supported by, for example, the tight
colour--luminosity relation of early-type galaxies found in galaxy
clusters \citep[e.g.,][]{bower92, ellis97}.

While different evolution in different models makes distant massive
galaxies a unique test-bed for galaxy formation scenarios, observations
have not yet provided evidence for the evolutionary path
of those galaxies. The major obstacle in observations originates
from the scarcity of galaxies at the high end of the galaxy mass
function; it means that not only it is hard to find the population
but also cosmic variance, the field-to-field variation of observed
volume density arising from large-scale structure, is significant. In
the last decade, many large programmes of optical-band imaging
have been carried out, providing excellent data sets with which
to investigate distant red old galaxies in wide fields of sky exceeding
a whole deg$^2$ 
\citep[e.g.,][]{bell04, borch06, cimatti06, willmer06, brown07, faber07}.
They consistently suggest that the total
stellar mass locked in red galaxies with luminosities around
$L \sim L^*$, where $L^*$ is the characteristic luminosity of the luminosity
function, has at least doubled since $z \sim 1$. Some of them also claim
little growth in the number of very luminous galaxies well above $L \sim L^*$. 
However, while luminous red galaxies roughly correspond
to massive galaxies, it is not clear how well the evolution in the
number of galaxies at the steep high end of the mass function is
understood from these results, since the much more numerous, less
massive galaxies with mass-to-luminosity ratios slightly less than
average could easily dominate the observed numbers of luminous
galaxies. The above authors also reveal that a field of view of the
order of a whole deg$^2$ is still not sufficient to conquer the uncertainty
arising from cosmic variance for the high-end populations of the
galaxy mass function.

The advent of the United Kingdom Infrared Telescope (UKIRT)
InfraredDeep Sky Survey \citep[UKIDSS;][]{lawrence07} provides
a unique opportunity to produce an ideal sample to trace various
aspects of massive galaxies in the distant Universe. Here we report
the results of a $K$-band survey with optical ($u$, $g$, $r$, $i$, $z$ band)
and near-infrared ($Y$, $J$, $H$ band) photometry and optical spectra,
focusing on massive ($M_{\star} > 10^{11} M_{\odot}$) galaxies out to $z = 1$ in an unprecedented
large area covering 55.2 deg$^2$. The $K$-band photometry
provides robust estimates of galaxy stellar masses \citep[e.g.,][]{matsuoka08} 
while the very large field of view significantly suppresses
cosmic variance, which allow us to conduct a unique analysis of the
mean properties of distant massive galaxies.

This paper is organized as follows. In Section 2 we describe the
data sources and reduction process to extract the galaxy sample from
the available data. Photometric redshifts and stellar masses are measured
for each galaxy in Section 3. In Section 4, the number-density
evolution of massive galaxies and the associated uncertainties are
explored. We then discuss the star-forming properties of galaxies
and the compatibility of the present results with previous measurements
in Section 5. A summary follows in Section 6. Throughout
this paper, we adopt the concordance cosmology of $H_0$ = 70 km s$^{-1}$ Mpc$^{-1}$, $\Omega_{\rm M} = 0.3$, 
and $\Omega_{\Lambda} = 0.7$. 
Magnitudes are expressed
in the Vega magnitude system for the UKIDSS near-infrared
bands and in the AB magnitude system for the optical bands.

\section{Data Sources and Reduction}

\subsection{Near-Infrared Photometry}

We extract from the Data Release 3 (DR3; Warren et al., in preparation)
of the UKIDSS/Large Area Survey (LAS) the $K$-band sources
with right ascensions from 1$^{\rm h}$ 15$^{\rm m}$ to 3$^{\rm h}$ 6$^{\rm m}$ 
on the Sloan Digital
Sky Survey \citep[SDSS;][]{york00} southern equatorial stripe (see
Section 2.2). The range in right ascension is chosen so that the $K$-band
observations are fairly complete within the sample area. More
than half of the $K$-band sources have also been observed in the $Y$,
$J$ and/or $H$ bands. We exclude the sources assigned with serious
quality flags corresponding to \texttt{ppErrBits} attributes larger than 31,
or near (within 20 arcsec of) the detector edges of any exposure. We
also exclude those sources near bright sources. This is achieved by
searching for bright point and extended sources in the Two-Micron
All-Sky Survey \citep[2MASS;][]{skrutskie06} catalogues and rejecting
all the LAS sources within sufficiently large distances of the
bright 2MASS sources. The total effective area of the observations
defining our sample is 55.2 deg$^2$.

We retrieved all the images in our sample area from the DR3 data
base, and used the \texttt{SOURCE EXTRACTOR}, version 2.5 \citep{bertin96}
for magnitude measurements. The total magnitudes ($m_{\rm tot}$) of
the sources are measured with the \texttt{SOURCE EXTRACTOR} total magnitude
algorithm \texttt{MAG\_AUTO}. We measure the aperture magnitudes
($m_{\rm ap}$) with circular apertures of several diameters, 4.5, 3.3, 2.8 and
2.6 arcsec, which correspond to 20 kpc at $z$ = 0.3, 0.5, 0.7 and 0.9,
respectively. Since the seeing condition is generally superior in the
UKIDSS LAS ($\sim$0.8 arcsec) to that in the optical observations by
the SDSS ($\sim$1.5 arcsec), the near-infrared images were smoothed
with the Gaussian kernel in such a way that the resultant full widths
at half-maximum (FWHMs) of stellar profiles are similar to those in
the optical images of the same field. The seeing measurements and
smoothing were performed in each of the small rectangular subareas
of approximately 9 $\times$ 13 arcsec$^2$. Then we ran the \texttt{SOURCE EXTRACTOR}
on the smoothed images to measure the aperture magnitudes of the
sources.

The detection completeness of the $K$-band sources can be estimated
by comparing the numbers of the LAS detections with those
of the much deeper UKIDSS Deep Extragalactic Survey \citep[DXS;][]{lawrence07}
in an overlapping field. This 0.6-deg$^2$ field is
a part of the DXS VIMOS 4 field, which is centred at RA 22$^{\rm h}$ 17$^{\rm m}$,
Dec. $+$00$^{\circ}$ 24' on the celestial equator. While the field is outside
the RA range of our LAS fields, we confirmed that the evaluation
sample of this field has a similar distribution of magnitudes and
their errors to those for our actual sample. Below we show that
the derived detection-completeness function reproduces the galaxy
number counts from our sample, in excellent agreement with previous
measurements, while it gives the lower limit of the detection
completeness for our massive galaxies. We define our sample as
being brighter than the limiting total magnitude $K_{\rm tot}$ = 17.9 mag,
where the detection completeness is higher than 0.5.

\subsection{Optical Photometry \label{subsec:optphot}}

We use the optical $u$, $g$, $r$, $i$ and $z$-band images on the SDSS southern
equatorial stripe. The stripe has been observed repeatedly in
the SDSS-II Supernova Survey \citep{frieman08} during 2005--
2007, as well as in the original SDSS. We retrieved all the available
images taken on the stripe from the SDSS Data Release 6 Supplemental
and Supernova Survey data bases \citep{adelman08}, and stacked them 
in each of the five bands. The images
observed in runs 2738 and 3325 are set apart from others, since they
are taken with the standard survey conditions of the original SDSS
and can work as the reference frames for stellar photometry. We
measure, for each retrieved image, the mean and root-mean-square
(rms) of the sky counts and the sky transparency at the observation
by comparing the stellar photometry of the relevant frame with that
of the reference frames. After discarding the worst 5 per cent of the
retrieved images with the largest rms of sky counts, which we find
is sufficient to reject apparently flawed exposures, the images are
zero-shifted and scaled according to the sky-count statistics and then
stacked by the inverse-of-variance weighted average using \texttt{IRAF}. The
photometric calibration of the stacked images is achieved by comparing
the stellar photometry with that of the reference frames. We
find that the stellar magnitudes on the stacked and reference frames
are in excellent agreement, with rms less than 0.05 mag (Fig. 1).
Fig. 2 shows the comparison of the original and stacked r-band
images of the same field. More than 100 original SDSS frames contribute
to each of the stacked frames, and the latter images are on
average $>$2mag deeper than the former images.

\begin{figure}
  \includegraphics[width=84mm]{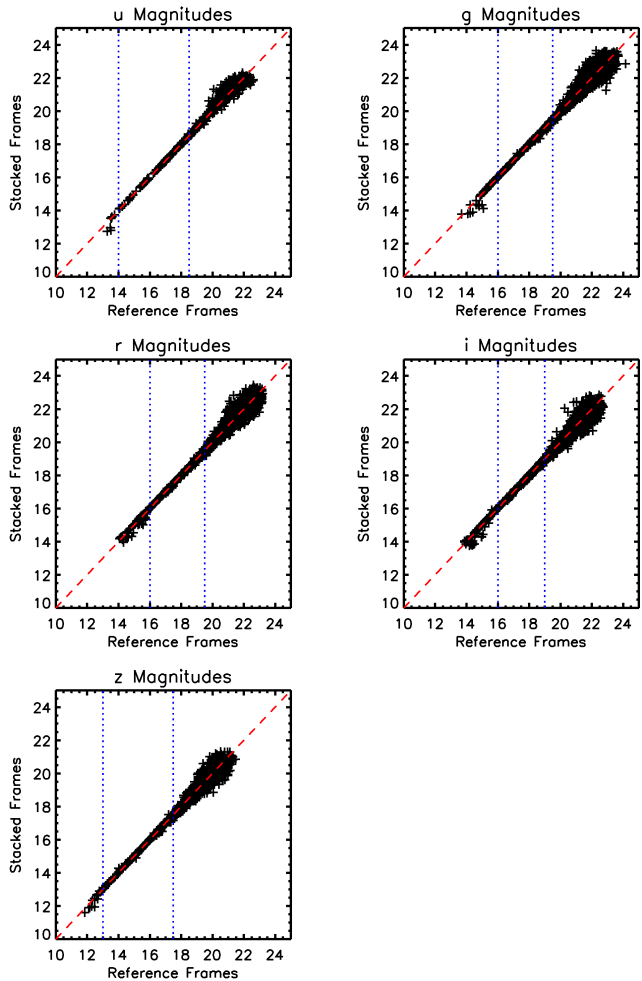}
  \caption{Comparisons of stellar magnitudes in the stacked and reference
frames in $u$ (top left), $g$ (top right), $r$ (middle left), $i$ (middle right) and $z$
(bottom left) bands. The dashed lines represent the locus where the two
measurements are identical. The rms errors of the photometric calibration
are calculated in the magnitude ranges shown by the dotted lines, where the
stellar photometry is most reliable.}
  \label{ref_vs_stack_phot}
\end{figure}

\begin{figure}
  \includegraphics[width=84mm]{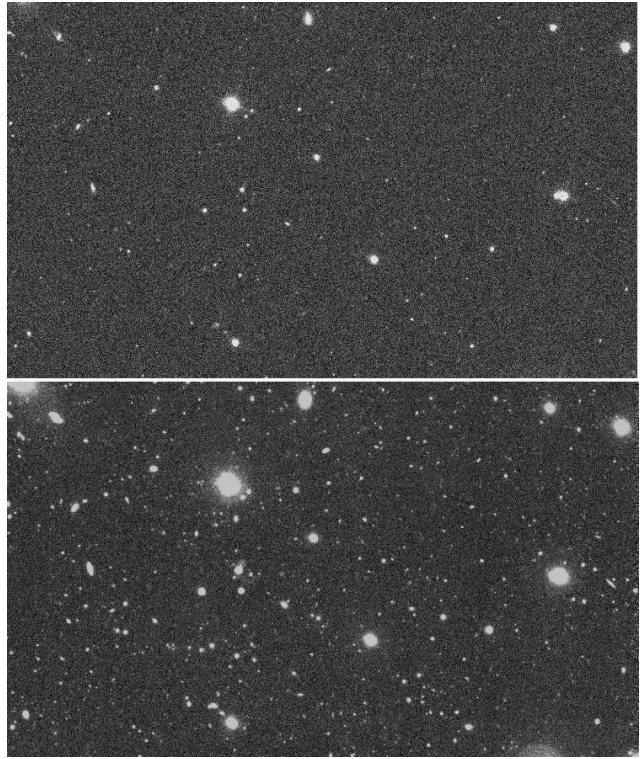}
  \caption{Original (top) and stacked (bottom) SDSS $r$-band images of the
same field.}
  \label{ref_vs_stack_image}
\end{figure}

We run the \texttt{SOURCE EXTRACTOR} on the stacked images to extract
detected sources. The groups with four or more pixels whose counts
are 1.5$\sigma$ above the local background level are identified as sources.
For every detection, we measure the aperture magnitudes in the
4.5-, 3.3-, 2.8- and 2.6-arcsec diameter apertures as we did for the
near-infrared images. The photometry errors are calculated from
the source photon counts and the background noise. The extracted
sources are cross-identified with the $K$-band sources within the
maximum paring tolerance of 1.0 arcsec. Owing to the deep stacking
of the SDSS images, nearly 90 per cent of the $K$-band sources have
counterparts in the $g$, $r$, $i$ and $z$ bands. Nearly 40 per cent of the
$K$-band sources also have counterparts in the $u$ band.

\subsection{Optical Spectroscopy}

We exploit the two redshift surveys carried out on the SDSS southern
equatorial stripe; the VIMOS-VLT Deep Survey \citep[VVDS;][]{lefevre05} 
and the DEEP2 Redshift Survey \citep{davis03}. 
Among the four fields of the VVDS 'Wide' survey, we use
the 4-deg$^2$ field of the F22 (2217$+$00), which lies on the celestial
equator, centred at RA 22$^{\rm h}$ 17$^{\rm m}$ 50$^{\rm s}$.4 and 
Dec. $+$00$^{\circ}$ 24$^{\rm m}$ 00$^{\rm s}$. This field
coincides with the UKIDSS DXS VIMOS 4, and is observed in both
the LAS and the DXS. We use Field 4 (RA 02$^{\rm h}$ 30$^{\rm m}$, Dec. $+$00$^{\circ}$ 00$^{\rm m}$) 
of the DEEP2, one of the '1-h survey' fields placed on our
LAS field. The VVDS adopts a pure $I$-band flux-limited selection
of the sample while the DEEP2 imposes strict colour pre-selection
on the spectroscopic targets to favour galaxies at $z > 0.7$; thus the
two surveys are complementary in terms of the sample selection.
We use only the spectroscopic sample with high redshift-quality
flags, \texttt{zflag}/\texttt{zQ} = 3 or 4 for the VVDS/DEEP2. As a result, we obtain
253 LAS $K$--VVDS and 375 LAS $K$--DEEP2 galaxies, as well
as 1084 DXS $K$--VVDS galaxies. We note that the redshift surveys
are deep enough that essentially all the LAS $K$ sources could be
sampled.

\subsection{Star/Galaxy Classification}

The $K$-band sources separate clearly into stars and galaxies on
the $r - z$ versus $z - K$ diagram as shown in Fig. 3. The colours
are measured with the 2.8-arcsec diameter aperture magnitudes.
We define the demarcation between stars and galaxies along the
minimum surface density on this diagram, i.e. the sources redder
than $z - K = 0.52 (r - z) + 1.74$ are classified as galaxies. The
additional criterion of $r - z < 4$ is set for galaxies to exclude cool
dwarf stars from the sample. We obtain 259 082 galaxies with these
classification criteria. The VVDS classification based on spectra
confirms that the above scheme works very well, yielding a rate
of misclassification (stars classified as galaxies and vice versa) of
less than $\sim$1 per cent. The $K$-band sources without $r$- and/or $z$-band
detections are excluded from the sample, since their extremely red
colours suggest that they are mostly galaxies beyond $z = 1$. Actually,
we find that more than 95 per cent of the LAS $K$--VVDS and LAS
$K$--DEEP2 galaxies are detected in both $r$ and $z$ bands in any redshift
bins at $z \le 1$.

\begin{figure}
  \includegraphics[width=84mm]{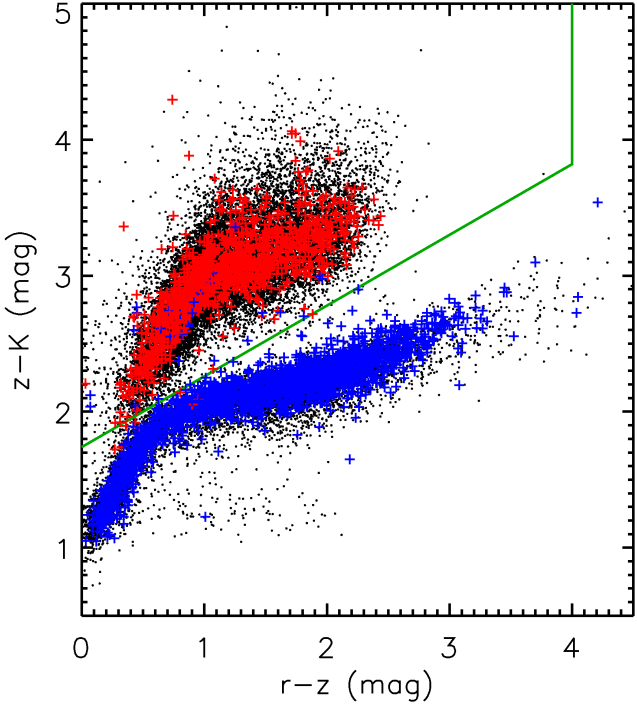}
  \caption{The $r - z$ versus $z - K$ diagram for a subset of $K$-band sources
(black), on which the VVDS stars (blue) and galaxies (red) are superimposed.
The green line represents the adopted star/galaxy classification
criteria.}
  \label{star_gal}
\end{figure}

We show the $K$-band differential number counts of the extracted
galaxies in Fig. 4. They are in excellent agreement with previous
measurements \citep{daddi00, huang01} down
to the limiting magnitude of $K_{\rm tot}$ = 17.9 mag after the detection-completeness
correction is applied. This suggests that we have successfully
constructed a well-defined sample of galaxies through the
above processes.

\begin{figure}
  \includegraphics[width=84mm]{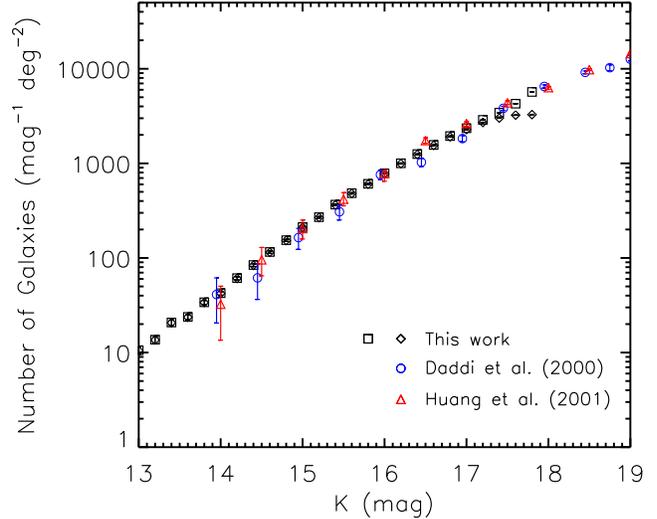}
  \caption{The $K$-band differential number counts of the extracted galaxies
with (squares) and without (diamonds) detection-completeness correction.
The error bars denote Poisson noise. The blue circles and the red triangles
represent the measurements of \citet{daddi00} and \citet{huang01},
respectively.}
  \label{num_counts}
\end{figure}

\section{Redshift and Stellar mass Measurements}

\subsection{Photometric Redshift \label{subsec:photoz}}

We estimate the redshifts of galaxies from the observed broad-band
colours, measured in the 2.8-arcsec diameter aperture, with
the optimized template-fitting method following \citet{ilbert06}.
We present a short summary of the method below, while the full
description of the concept and details can be found in the above
reference.

We choose the DXS $K$--VVDS galaxies to optimize the spectral
templates, leaving the LAS $K$--VVDS and LAS $K$--DEEP2 galaxies
as the evaluation sets of the redshift measurements (Table 1). First,
the DXS $K$--VVDS galaxies are classified into five spectral types,
Ell, Sbc, Scd, Irr and starburst (SB) by least-$\chi^2$ fitting with the
appropriate amounts of dust extinction assuming the extinction laws
of the Small Magellanic Cloud (SMC) by \citet{pei92} for Scd and Irr
and that of starburst galaxies by \citet{calzetti00} for SB. The
initial spectral templates are taken from \citet{cww80} for Ell, Sbc, Scd 
and Irr and from \citet{kinney96} for
SB. Then, for each filter $f$, we minimize the sum
\begin{equation}
\psi^2 = \sum_{\rm galaxy}^{} 
\biggl( \frac{F_{\rm obs}^f - A \times F_{\rm model}^f - s^f}{\sigma_{\rm obs}^f} \biggr)^2,
\end{equation}
where $F_{\rm obs}^f$ and $\sigma_{\rm obs}^f$ are the observed flux and its error in the filter $f$.
The sum is taken over all the sample galaxies. The parameters 
$F_{\rm model}^f$ and $A$
represent the best-fitting template flux and its normalization
factor taken from the initial least-$\chi^2$ fitting. The last term $s^f$ is
a free parameter. While this term should be zero in the case of
a completely random uncertainty in the photometry, we find that
it has a non-zero value in every filter. These values are at most
0.05 mag and are comparable with the expected uncertainty in the
photometric zero-point calibration.

\begin{table}
 \caption{Summary of the spectroscopic sample.}
 \label{tab:spec_sample}
 \begin{tabular}{@{}lcc}
  \hline
  Sample   & Number & Use$^{*1}$\\
  \hline
  DXS $K$ -- VVDS & 1084 & training\\
  LAS $K$ -- VVDS$^{*2}$ &  253 & evaluation\\
  LAS $K$ -- DEEP2 & 375 & evaluation\\
  \hline
 \end{tabular}

 \medskip
 $^{*1}$Use in the photometric-redshift measurements.\\
 $^{*2}$Approximately 70\% of the LAS $K$ -- VVDS galaxies are also the members of the DXS $K$ -- VVDS galaxies.
\end{table}

The DXS $K$--VVDS galaxies are re-classified into five spectral-type
groups with the terms $s^f$ considered. In each group, the observed
broad-band fluxes of galaxies are converted to the rest frame
according to the spectroscopic redshifts, after being normalized and
de-reddened by the best-fitting normalization factor $A$ and the dust
extinction. Since the galaxies have various redshifts, this conversion
generates a continuous spectral energy distribution for each
spectral type of galaxies over the relevant rest-frame wavelength
range. We sort the rest-frame fluxes according to their wavelengths
and bin them by groups of points, and connect the median flux in
each bin to produce the optimized templates. We keep the extrapolations
provided by the initial templates in ultraviolet and infrared
wavelengths where no broad-band data are available. The starburst
template is not optimized, in order to retain the emission lines in
the template. Finally, these optimized templates are interpolated to
produce a total of 62 templates, the first being Ell and the last being
SB, to improve the sampling of the redshift--colour space. Below
we define the spectral type of each galaxy using the best fits from
among these 62 templates.

The created spectral templates are fitted to the observed colours
of the actual sample to measure their redshifts. We evaluate the
measurement accuracy by applying the same procedure to the LAS
$K$--VVDS and LAS $K$--DEEP2 galaxies, as well as the DXS $K$--VVDS 
galaxies. The DXS $K$--VVDS galaxies are reduced in number
according to the LAS $K$-band detection completeness and are
given additional random photometry errors in order to simulate the
LAS $K$-band galaxies.We find that the photometric redshifts ($z_{\rm phot}$)
are well correlated with the spectroscopic redshifts ($z_{\rm spec}$) as shown
in Fig. 5, thanks to our wide and relatively fine wavelength coverage
in the $u$ through $K$ bands. The deviation between the two (photometric
and spectroscopic) measurements closely follows a Gaussian
distribution with standard deviation $\sigma_{{\Delta}z/(1+z)} \sim 0.04$ 
(${\Delta}z = z_{\rm spec} - z_{\rm phot}$) for all three sets of the evaluation sample. Note that
$\sim$70 per cent of the LAS $K$--VVDS galaxies are also members of
the DXS $K$--VVDS galaxies and account for about a quarter of the
sample used to build up the spectral templates, so that the LAS
$K$--VVDS galaxies do not provide a completely independent test of
the photometric-redshift accuracy. As a further test, we created another
template set from the DXS $K$--VVDS galaxies omitting these
LAS $K$--VVDS galaxies and repeated the photometric-redshift measurement.
This test again gives $\sigma_{{\Delta}z/(1+z)} \sim 0.04$, which indicates that
it is a robust estimate of the photometric-redshift uncertainty. We
show the uncertainty as a function of redshift and stellar mass (as
determined below) in Table 2. They are relatively large in the lowest
and highest redshift bins for larger stellar mass classes, for which
relatively small numbers of sample contribute to the spectral templates.
We also show the uncertainty as a function of spectral type
in Table 3, which suggests there is little variation of uncertainty
among the different spectral types.

\begin{figure}
  \includegraphics[width=84mm]{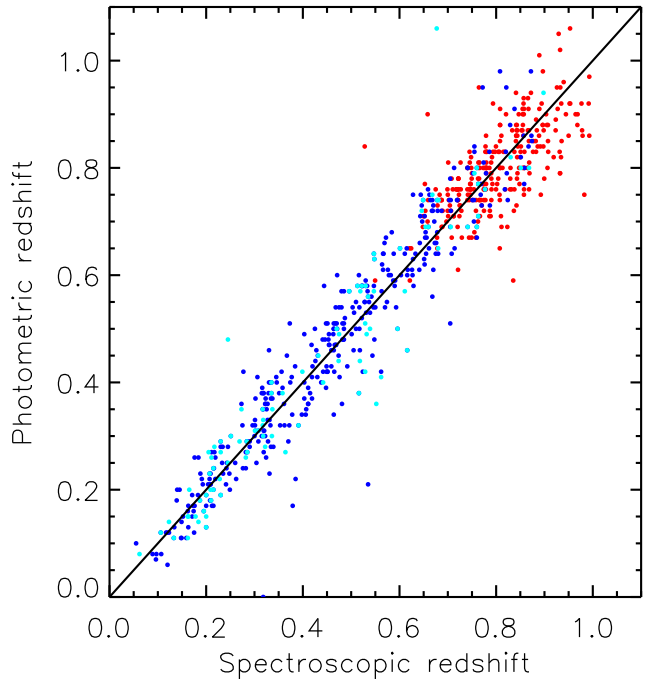}
  \caption{Comparison of the photometric and spectroscopic redshift measurements
of LAS $K$--VVDS (light blue), DXS $K$--VVDS (blue) and LAS
$K$--DEEP2 (red) galaxies. The solid line shows the locus where the two
measurements are identical.}
  \label{photz_vs_specz}
\end{figure}

\begin{table}
 \caption{Uncertainty ($\sigma_{{\Delta}z/(1+z)}$) in photometric redshift.}
 \label{tab:photo-z_err}
 \begin{tabular}{@{}lcccc}
  \hline
             &            & log $M_{\star}$         &           \\
  Redshift   & 10.0 -- 10.5 & 10.5 -- 11.0 & 11.0 -- 11.5 & 11.5 -- 12.0\\
  \hline
  0.2 -- 0.4 & 0.037 (50) & 0.046 (51) & 0.043 ( 6) & --- ( 0) \\
  0.4 -- 0.6 & 0.040 (35) & 0.037 (55) & 0.039 (25) & --- ( 0) \\
  0.6 -- 0.8 & 0.021 (35) & 0.030 (73) & 0.032 (93) & 0.034 (20) \\
  0.8 -- 1.0 & 0.033 (17) & 0.038 (71) & 0.046 (55) & 0.047 ( 7) \\
  \hline
 \end{tabular}

 \medskip
 Note --- Numbers in parentheses represent the number in the evaluation samples.
\end{table}

\begin{table}
 \caption{Uncertainty ($\sigma_{{\Delta}z/(1+z)}$) in photometric redshift.}
 \label{tab:photo-z_err2}
 \begin{tabular}{@{}cc}
  \hline
  Spectral type & $\sigma_{{\Delta}z/(1+z)}$\\
  \hline
  Ell -- Sbc & 0.035 (386)\\
  Sbc -- Scd & 0.037 (171)\\
  Scd -- Irr & 0.036 ( 35)\\
  Irr -- SB  &  ---  (  0)\\
  \hline
 \end{tabular}

 \medskip
 Note --- Numbers in parentheses represent the number in the evaluation samples.
\end{table}

\subsection{Stellar Mass \label{subsec:stellarmass}}

The stellar masses ($M_{\star}$) of galaxies are determined by fitting to the
observed colours the stellar population synthesis models of \citet{bc03}. 
The aperture magnitudes ($m_{\rm ap}$) of the 4.5-, 3.3-,
2.8- and 2.6-arcsec apertures are used for the fitting of galaxies with
photometric redshifts $z$ = 0.2--0.4, 0.4--0.6, 0.6--0.8, and 0.8--1.0,
respectively, so that we sample the stellar populations consistently
within the central $\sim$20 kpc of all the galaxies. The resultant stellar
mass is then scaled by $10^{0.4 (K_{\rm ap}-K_{\rm tot})}$ to correct for aperture loss.
We adopt the standard configurations with the Padova 1994 stellar
evolutionary tracks and BaSel 3.1 spectral library for the \citet{bc03} models. 
We assume three values of metallicity: 0.2,
1 and 2.5 $Z_{\odot}$, where $Z_{\odot}$ is the solar metallicity. The star-formation
history is assumed to take the exponentially declining form 
$\tau^{-1}$ exp($-t/\tau$), where the $e$-folding time $\tau$ is a free parameter, with
the \citet{salpeter55} initial mass function (IMF). Our stellar mass
estimates can be approximately converted to those with another
commonly used IMF, that of \citet{chabrier03}, by adding $\sim$0.25 dex.
Other free parameters are the age $t$ of the stellar population and the
colour excess $E(B-V)$ due to the dust extinction of the stellar
radiation, following the SMC extinction curve of \citet{pei92}. These
parameters are varied over the plausible ranges of 
10 Myr $\le \tau \le$ 10 Gyr ($\Delta$log $\tau$ = 0.2), 
10 Myr $\le t \le$ 10 Gyr ($\Delta$log $t$ = 0.1), and 
$0.0 \le E(B-V) \le 0.5$ mag ($\Delta$$E(B-V)$ = 0.05), 
and the best-fitting
parameter set is searched for by the least-$\chi^2$ method for
each galaxy (the values in parentheses represent the grid intervals).
The additional error of 0.05 mag is added in quadrature to all band
magnitudes in the fitting in order to take into account the uncertainty
in the photometric zero-point calibration.

We derive two kinds of stellar mass for the spectroscopic sample,
i.e. the stellar mass with spectroscopic redshifts ($M_{\star, spec}$) and
the stellar mass with photometric redshifts ($M_{\star, phot}$). The difference
between the two measures, $\Delta {\rm log} M_{\star} = {\rm log} M_{\star, spec} - {\rm log} M_{\star, phot}$, is
found to be clearly correlated with the photometric redshift deviation
${\Delta}z = z_{\rm spec} - z_{\rm phot}$. Such a correlation is expected, since
a larger $z_{\rm phot}$ leads to a larger estimate of the galaxy luminosity,
which then leads to a larger estimate of stellar mass $M_{\star, phot}$.
The observed relation between ${\Delta}z$ and $\Delta {\rm log} M_{\star}$ is actually quite
consistent with this expected correlation. Another expected cause
of the correlation between ${\Delta}z$ and $\Delta {\rm log} M_{\star}$ comes from the fact
that larger estimates of $z_{\rm phot}$ lead to systematically shorter rest-frame
wavelengths to which each of the observing wavebands
corresponds. After removal of the above first component of the
systematic correlation, we found marginal evidence for the second
correlation in our sample, which is $\Delta {\rm log} M_{\star}$ = $+/-$ ($0.04 \pm 0.07$) 
when ${\Delta}z$ is negative/positive. In addition, we consider the uncertainty
associated with the least-$\chi^2$ model fitting. It is evaluated
by the 1$\sigma$ confidence surface of the $\chi^2$ distributions in the fitting
parameter space.

We show the total amplitudes of the stellar mass uncertainty
($\sigma_{\Delta {\rm log} M_{\star}}$) as a function of redshift and stellar mass in Table 4. Those
as a function of spectral type are shown in Table 5. The above estimates
of error amplitudes and the correlations between ${\Delta}z$ and $\Delta {\rm log} M_{\star}$
are taken into account in the Monte Carlo simulation presented
below. Note that further different assumptions on the stellar
mass estimation, such as different stellar population synthesis
models and different IMF, can cause additional uncertainty in the
derived properties of galaxies. We will address this issue in Section
5.

\begin{table}
 \caption{Uncertainty ($\sigma_{\Delta {\rm log} M_{\star}}$) in stellar mass.}
 \label{tab:logM_err}
 \begin{tabular}{@{}lcccc}
  \hline
             &            & log $M_{\star}$         &           \\
  Redshift   & 10.0 -- 10.5 & 10.5 -- 11.0 & 11.0 -- 11.5 & 11.5 -- 12.0\\
  \hline
  0.2 -- 0.4 & 0.19 (50) & 0.20 (51) & 0.22 ( 6) & --- ( 0) \\
  0.4 -- 0.6 & 0.19 (35) & 0.17 (55) & 0.16 (25) & --- ( 0) \\
  0.6 -- 0.8 & 0.12 (35) & 0.16 (73) & 0.17 (93) & 0.19 (20) \\
  0.8 -- 1.0 & 0.14 (17) & 0.17 (71) & 0.19 (55) & 0.25 ( 7) \\
  \hline
 \end{tabular}
 \medskip
 Note --- Numbers in parentheses represent the number in the evaluation samples.
\end{table}

\begin{table}
 \caption{Uncertainty ($\sigma_{\Delta {\rm log} M_{\star}}$) in stellar mass}
 \label{tab:logM_err2}
 \begin{tabular}{@{}cc}
  \hline
  Spectral type  & $\sigma_{\Delta {\rm log} M_{\star}}$\\
  \hline
  Ell -- Sbc & 0.16 (386)\\
  Sbc -- Scd & 0.17 (171)\\
  Scd -- Irr & 0.17 ( 35)\\
  Irr -- SB  &  --- (  0)\\
  \hline
 \end{tabular}

 \medskip
 Note --- Numbers in parentheses represent the number in the evaluation samples.
\end{table}

\section{Results}

We define our massive galaxy sample using two stellar mass classes,
i.e. $10^{11.0-11.5} M_{\odot}$ galaxies with 
$10^{11.0} M_{\odot} < M_{\star} < 10^{11.5} M_{\odot}$ and $10^{11.5-12.0} M_{\odot}$ galaxies with 
$10^{11.5} M_{\odot} < M_{\star} < 10^{12.0} M_{\odot}$. 
The
galaxies are grouped into four redshift bins, $z$ = 0.2 -- 0.4, 0.4 -- 0.6, 0.6 -- 0.8, and 0.8 -- 1.0.
Total numbers included in the sample are
summarized in Table 6. The median photometry errors in the $r$-, $z$-, and
$K$-band aperture magnitudes and in the $K$-band total magnitudes
($r_{\rm ap}$, $z_{\rm ap}$, $K_{\rm ap}$, $K_{\rm tot}$) are also listed. 
The aperture magnitude errors
are generally smaller than the typical uncertainty in the photometric
zero-point calibration ($\sim$0.05 mag).

\begin{table*}
 \caption{Summary of the massive galaxy sample.}
 \label{tab:sample_num}
 \begin{tabular}{@{}cclcl}
  \hline
           &  log $M_{\star}$ = &  11.0 -- 11.5 & log $M_{\star}$ = &  11.5 -- 12.0\\
  Redshift &   Number &  Photometry Error$^*$              &   Number   &  Photometry Error$^*$ \\
  \hline
  0.2 -- 0.4 &  9,720 & ($<$0.01, $<$0.01, 0.01, 0.05) & 1,408 &  ($<$0.01, $<$0.01, $<$0.01, 0.03)\\
  0.4 -- 0.6 & 15,300 & ($<$0.01, $<$0.01, 0.02, 0.08) &   572 &  ($<$0.01, $<$0.01, 0.01, 0.04)\\
  0.6 -- 0.8 & 18,582 & (0.02, 0.02, 0.03, 0.13)       &   815 &  (0.01, 0.01, 0.02, 0.09)\\
  0.8 -- 1.0 & 12,371 & (0.04, 0.02, 0.04, 0.16)       &   613 &  (0.03, 0.02, 0.03, 0.12)\\
  \hline
 \end{tabular}

 \medskip
 Note ($^*$) --- Median photometry errors in the r-, z- and K-band aperture magnitudes and in the K-band total magnitudes
 ($r_{\rm ap}$, $z_{\rm ap}$, $K_{\rm ap}$, $K_{\rm tot}$).
\end{table*}

We show the differential number counts of the massive galaxies
in Fig. 6. It shows that the number-count distributions of
the $10^{11.5-12.0} M_{\odot}$ galaxies have faint-end drop-offs at magnitudes
brighter than the limiting magnitude, which assures the near-complete
detection of this population. On the other hand, the faintest
of the $10^{11.0-11.5} M_{\odot}$ galaxies at high redshifts ($z > 0.6$) fall below
the limiting magnitude, and are thus left uncounted. In order
to estimate the lost fraction of $10^{11.0-11.5} M_{\odot}$ galaxies at these redshifts,
we derive the detection completeness specifically for these
galaxies as follows. We take each of the $10^{11.0-11.5} M_{\odot}$ galaxies in
the lowest redshift bin and assign random redshifts in the $0.6 < z < 0.8$ and 
$0.8 < z < 1.0$ ranges. The galaxies are dimmed and
reduced in apparent size according to the assigned redshifts, placed
on random positions of the LAS $K$-band images, and then extracted
by \texttt{SOURCE EXTRACTOR} in the same way as the actual sample sources
are detected. The recovery rate of the embedded objects as a function
of their magnitudes gives the detection completeness of the
galaxies, which we find is significantly better than that of the whole
sample derived before. The 50 per cent detection completeness is
actually achieved at $K_{\rm tot}$ = 18.6 mag instead of $K_{\rm tot}$ = 17.9 mag,
and almost all galaxies brighter than $K_{\rm tot}$ = 17.9 mag are detected.
With the new detection-completeness function taken into account,
the fractions of $10^{11.0-11.5} M_{\odot}$ galaxies fainter than the formal limiting
magnitude ($K$ = 17.9 mag, thus uncounted) are 11 and 16 per
cent at $z$ = 0.6--0.8 and 0.8--1.0, respectively

\begin{figure}
  \includegraphics[width=84mm]{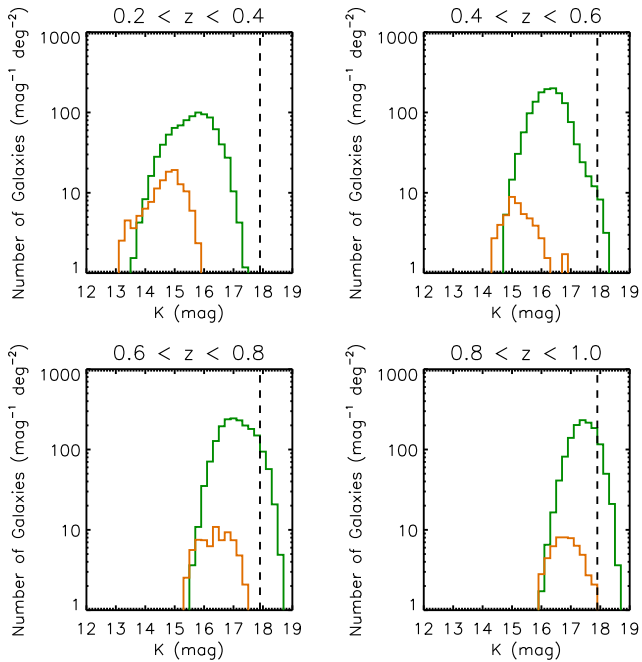}
  \caption{The differential number counts of $10^{11.0-11.5} M_{\odot}$ (green) and
$10^{11.5-12.0} M_{\odot}$ (orange) galaxies at $z$ = 0.2--0.4 (top left), 0.4--0.6 (top
right), 0.6--0.8 (bottom left) and 0.8--1.0 (bottom right). The black dashed
lines represent our formal limiting magnitude of $K_{\rm tot}$  = 17.9 mag.}
  \label{z_num2_4}
\end{figure}

We estimate the uncertainty in the measured numbers of massive
galaxies by a Monte Carlo simulation, as follows. First, we generate
a mock galaxy catalogue containing 10 galaxies for each stellar
mass ($\Delta$log$M_{\star} = 0.1$) and redshift ($\Delta{z} = 0.02$) bin in the ranges
$8.0 <$ log $M_{\star}$ $< 13.0$ and $0.0 < z < 1.4$, where the numbers in
parentheses represent the bin widths. Each galaxy is assigned a
weighting factor corresponding to the number density of galaxies,
following the galaxy mass function of \citet{cole01}. Next the
redshifts of galaxies are given perturbations following the measured
uncertainty of the actual sample. The stellar masses are also given
perturbations, correlated with the photometric redshift perturbations
as explored in Section 3.2. Then the mock galaxies are weighted by
their weighting factors and redistributed, and counted in the redshift
and stellar mass bins to obtain the output mass function. We repeat
the calculation 100 times, varying the random components.

Fig. 7 shows the results of the simulation. We observe both systematic
and random components in the resultant error estimates.
The systematic component is evident in $M_{\star} > 10^{12.0} M_{\odot}$ bins. This
is the so-called Eddington bias \citep{eddington13}, caused by the
steep slope of the high end of the mass function; simply put, a small
portion of the less massive, much more numerous galaxies could
contaminate the more massive classes owing to measurement errors,
which significantly alters the steep part of the mass function.
This is why we limit our massive galaxy sample to those with $M_{\star} < 10^{12.0} M_{\odot}$; 
the systematic increases in number are found to be insignificant
for our mass ranges, i.e. negligible for the $10^{11.0-11.5} M_{\odot}$
galaxies and 40--60 per cent for the $10^{11.5-12.0} M_{\odot}$ galaxies. The
output number densities from the 100 repeated calculations scatter
around the systematic components, which yields the random
components of the measurement error. The standard deviations of
the scatter are $\sim$5 per cent and $\sim$12 per cent of the numbers of
$10^{11.0-11.5} M_{\odot}$ and $10^{11.5-12.0} M_{\odot}$ galaxies, respectively. Note that
the current estimate is not perfect, since we assume a non-evolving
galaxy stellar mass function at $0 < z < 1$, while our results show
clear signs of its evolution (see below). We will discuss this issue
further in the following section.

\begin{figure}
  \includegraphics[width=84mm]{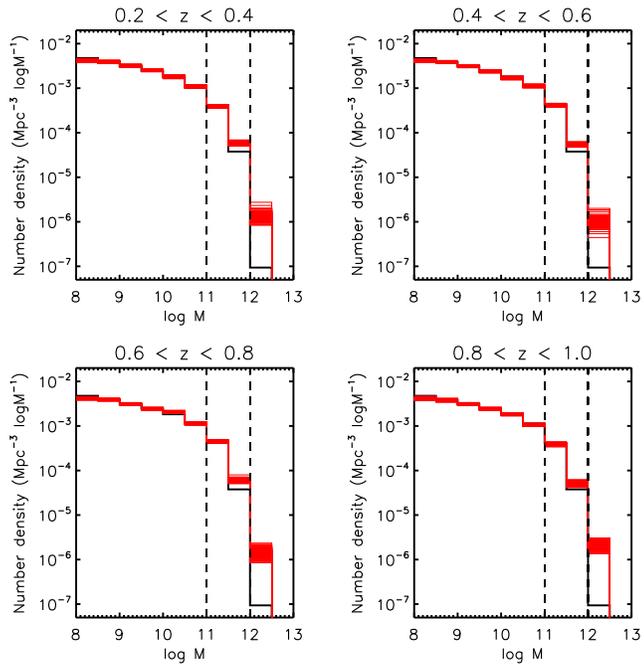}
  \caption{The input (black line) and output (100 red lines) galaxy mass
functions of the Monte Carlo simulation at $z$ = 0.2--0.4 (top left), 0.4--0.6
(top right), 0.6--0.8 (bottom left) and 0.8--1.0 (bottom right). The dashed
lines show the mass ranges in which our massive galaxy sample is defined.}
  \label{montecarlo}
\end{figure}

Another source of uncertainty comes from cosmic variance. We
estimate cosmic variance by dividing our sample into five subfields
along the right ascension, $\sim$11 deg$^2$ each, and then calculating
the fractional variation of the measured numbers of massive
galaxies in these subfields. We find that the fractional variations
are $<$10 per cent and $<$20 per cent for the $10^{11.0-11.5} M_{\odot}$ and $10^{11.5-12.0} M_{\odot}$  
galaxies, respectively, in each of the four redshift
bins. Considering that the total fields are five times larger than the
subfields, we conclude that cosmic variance could affect the measured
number of massive galaxies in each redshift bin by up to 5 per
cent and 10 per cent for the two classes of galaxies, respectively.
The above estimates are roughly consistent with the theoretical
predictions provided by \citet{somerville04}.

We show our results with regard to the number-density measurements
in Table 7, and plot them along with the measurements for
the local Universe \citep{cole01} in Fig. 8. The error bars take
into account all uncertainties considered above, as well as the Poisson
noises. The number densities of the $10^{11.5-12.0} M_{\odot}$ galaxies
are corrected for Eddington bias, although this correction has little
significance for our final conclusions. The local densities were normalized
to take into account the assumptions of \citet{cole01}
with regard to stellar population synthesis that differ from ours; the
major difference is that they assume a constant formation redshift
of galaxies at $z_f = 20$, while we vary the age of the stellar population
as a free parameter. The normalization factor, $\sim0.4$, is derived
by applying their assumption to our $10^{11.0-11.5} M_{\odot}$ galaxies in the
lowest redshift bin $z$ = 0.2--0.4. As seen in the figure, we find that
the most massive ($10^{11.5-12.0} M_{\odot}$) galaxies have experienced rapid
evolution in number since $z = 1$. On the other hand, the number
densities of the less massive ($10^{11.0-11.5} M_{\odot}$) systems show a rather
mild evolution during the same period.

\begin{table}
 \caption{Number densities of the massive galaxies.}
 \label{tab:number_densities}
 \begin{tabular}{@{}lcc}
  \hline
             &    log $M_{\star}$         &           \\
  Redshift   & 11.0 -- 11.5               & 11.5 -- 12.0\\
  \hline
  0.2 -- 0.4 & 10.9 $\pm$ 0.8      & 0.98 $\pm$ 0.15 \\
  0.4 -- 0.6 & 7.8  $\pm$ 0.6      & 0.19 $\pm$ 0.03 \\
  0.6 -- 0.8 & 6.8  $\pm$ 0.5      & 0.17 $\pm$ 0.03 \\
  0.8 -- 1.0 & 3.7  $\pm$ 0.3      & 0.11 $\pm$ 0.02 \\
  \hline
 \end{tabular}
 \medskip
\begin{flushleft}
 Note --- Number densities are given in units of $10^{-4}$ Mpc$^{-3}$ log $M_{\star}^{-1}$.
\end{flushleft}
\end{table}

\begin{figure}
  \includegraphics[width=84mm]{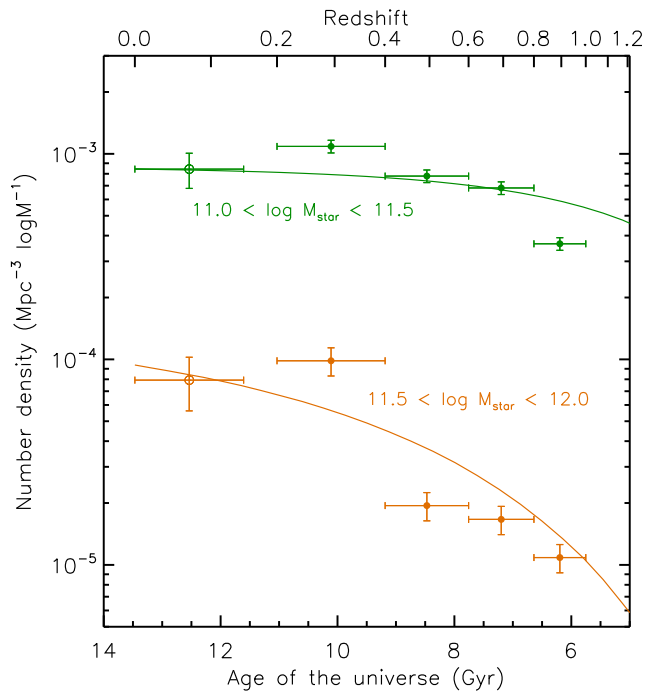}
  \caption{The number densities of massive galaxies versus the age of the
Universe or redshift. The green and orange symbols and lines represent the
$10^{11.0-11.5} M_{\odot}$ and $10^{11.5-12.0} M_{\odot}$ galaxies, respectively. The filled circles
represent our measurements while the open circles represent measurements
in the local Universe \citep{cole01}. The solid lines represent the predictions
of the Millennium Simulation with the semi-analytic galaxy formation
model (see text).}
  \label{number_densities}
\end{figure}

\section{Discussion \label{sec:discuss}}

The measured number density evolution of massive galaxies shows
clear signs of the hierarchical evolution of these systems. Such a
galaxy evolution scenario is predicted in the latest galaxy formation
models based on $\Lambda$CDM theory. In Fig. 8 we overlay the predictions
of the Millennium Simulation (Lemson et al. 2006), the largest
numerical simulation to date based on $\Lambda$CDM theory, with the semianalytic
galaxy formation model of \citet{delucia07},
scaled to fit to the local observations (by a factor of 0.4). The
stellar mass of the Millennium model has been shifted by $+$0.25 dex
in order to correct for the different IMFs adopted (the model adopts
the \citet{chabrier03} IMF). The observed hierarchical pattern of
evolution is consistent with the prediction of the model, while we
find some discrepancies between the observation and the model
(e.g. the $10^{11.0-11.5} M_{\odot}$ galaxies at $z$ = 0.8--1.0). 
This indicates that
the basic idea of the bottom-up construction of galaxy systems is
valid at least for the most massive galaxies with $M_{\star} > 10^{11} M_{\odot}$.

Meanwhile, we point out that our measurements plotted in Fig. 8
seem to have an apparently unnatural feature at $z$ = 0.2--0.4, where
the number densities are significantly high relative to the overall
trend considering their estimated errors, implying residual systematic
effects from the Eddington bias. We check this by investigating
the systematic increases in the number of massive galaxies in the
spectroscopic sample, from those obtained with the spectroscopic
redshifts to those with photometric redshifts (we consider only those
redshift and stellar mass classes with more than five spectroscopic
sources). The above Monte Carlo simulation shows that the use of
photometric redshifts (and the consequent stellar mass fluctuations)
is the dominant source of the Eddington bias. We list the measured
systematic increases in Table 8 separately for the VVDS and DEEP2
galaxies, since the amplitudes of the Eddington bias are subject to
the redshift distribution of sources. The DXS $K$--VVDS galaxies
have been given additional photometry errors and detection incompleteness
in order to simulate the LAS $K$ galaxies. The systematic
increases estimated in the Monte Carlo simulation are also listed in
Table 8. The table shows that the systematic increases measured in
the spectroscopic sample are mostly less than the estimates from the
Monte Carlo simulation, while the one for the VVDS $10^{11.0-11.5} M_{\odot}$ galaxies at $z = 0.2 - 0.4$ 
is exceptionally high. While it is based on
a small sample, it provides marginal evidence that the Eddington
bias is unexpectedly significant in this lowest redshift bin, due to
some unquantified sources of systematic uncertainty. In that case,
the number densities measured at $z$ = 0.2--0.4 should be regarded
as the upper limits.

We also note that we adopt a non-evolving galaxy stellar mass
function in the Monte Carlo simulation, while our results suggest
the steepening of the high end of the mass function toward high
redshifts; thus the estimated amount of Eddington bias should be
regarded as a lower limit. When such a steepening of the mass function
is taken into account in the correction of the Eddington bias, we
obtain even lower numbers of galaxies in the more massive classes
at higher redshifts, which would further strengthen our conclusion.

\begin{table*}
 \caption{Estimates of the Eddington bias$^{*1}$.}
 \label{tab:eddington_bias}
 \begin{tabular}{@{}ccccccc}
  \hline
           &  log $M_{\star}$ = &  11.0 -- 11.5 &        & log $M_{\star}$ = &  11.5 -- 12.0 &  \\
  Redshift &   MCS$^{*2}$  & VVDS$^{*3}$  & DEEP2$^{*3}$ &  MCS$^{*2}$ & VVDS$^{*3}$ & DEEP2$^{*3}$ \\
  \hline
  0.2 -- 0.4 & $<$ 0.1     & 0.7 (6)      & - (0)             &  0.6        &   - (0)  & - (0)    \\
  0.4 -- 0.6 & $<$ 0.1     & $<$ 0.1 (23) & - (2)             &  0.5        &   - (0)  & - (0)    \\
  0.6 -- 0.8 & 0.1         & 0.1 (21)     & $<$ 0.1 (72)      &  0.6        &  0.3 (7) & 0.3 (13)  \\
  0.8 -- 1.0 & $<$ 0.1     & - (2)        & $<$ 0.1 (53)      &  0.4        &   - (0)  & 0.1 (7)  \\
  \hline
 \end{tabular}
 \medskip
\begin{flushleft}
 $^{*1}$Rates of increase (increased amounts divided by the original values) of the number densities are listed.\\
 $^{*2}$Estimates from the Monte Carlo simulation.\\
 $^{*3}$Estimates from the VVDS and DEEP2 spectroscopic samples (the number in the samples is shown in the parentheses).\\
  \end{flushleft}
\end{table*}

Recently, \citet{marchesini09} provided a detailed analysis
of random and systematic uncertainties affecting the galaxy stellar
mass function. They adopt 14 different stellar mass estimations
with different combinations of metallicity, dust extinction law, stellar
population synthesis models and IMF, and show that the derived
number density of galaxies in a given stellar mass bin could be
altered by up to 1 dex. The 'bottom-light' IMFs in particular, with
a deficit of low-mass stars relative to a standard \citet{chabrier03}
IMF, give significantly different results for the stellar mass function
from other classical IMFs. A more top-heavy IMF at higher
redshifts is actually suggested by, for example, \citet{dave08} and
\citet{vandokkum08}. Thus we point out that our results are subject
to a systematic change of these stellar-population properties during
$0 < z< 1$. Future improvements in stellar population models based
on new observations are eagerly awaited to overcome these large
uncertainties inherent in stellar mass measurements.

In order to probe the star-forming properties of massive galaxies,
here we investigate their rest-frame optical colours. Nearby galaxies
are known to show a clear bimodality in the optical colour distribution,
in which early-type galaxies form a narrow red sequence that
is separated from blue star-forming populations by a valley of the
galaxy distribution \citep[e.g.,][]{strateva01, hogg03}. A
similar bimodality is observed out to $z > 1$ \citep[e.g.,][]{lin99, im02, bell04}. 
We calculate the rest-frame $U$- and
$V$-band magnitudes of massive galaxies by $k$-correcting the
nearest observed $r$, $i$ or $z$-band magnitudes, where the amounts of
the $k$-corrections are estimated from the best-fitting spectral population
synthesis models. We show the resultant colour distributions
in Fig. 9.

\begin{figure}
  \includegraphics[width=84mm]{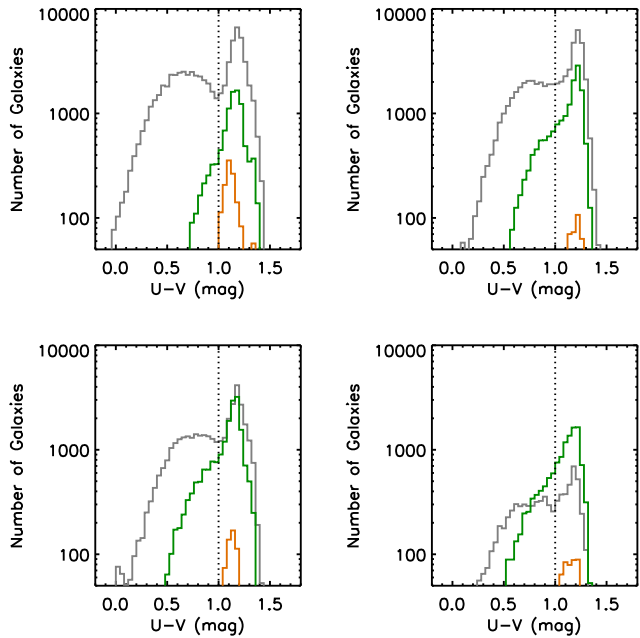}
  \caption{$U - V$ colour distributions of the massive galaxies at $z$ =
0.2--0.4 (top left), 0.4--0.6 (top right), 0.6--0.8 (bottom left) and 0.8--1.0 (bottom right). 
The grey, green and orange lines represent the
$< 10^{11.0} M_{\odot}$, $10^{11.0-11.5} M_{\odot}$ and $10^{11.5-12.0} M_{\odot}$ galaxies, respectively.
The dotted lines show the demarcation between the blue and red populations.}
  \label{cmd}
\end{figure}

As Fig. 9 shows, the less massive ($< 10^{11.0} M_{\odot}$) galaxies show
a clear colour bimodality, as expected, with the peak colour of the
red sequence at $U - V \sim 1.2$ and a valley of galaxy distributions
at $U - V \sim 1.0$ in all redshift bins. Compared with these
galaxies, massive ($> 10^{11.0} M_{\odot}$) galaxies are apparently dominated
by the red population, with conspicuous peaks at $U - V \sim 1.2$.
We divide the galaxies into blue and red populations at $U - V = 1.0$ 
and calculate the blue fractions (fractions of the blue population).
The associated errors are estimated by repeatedly giving
random fluctuations to the $U - V$ colours, taking into account the
uncertainties in the photometry and the amount of $k$-correction, and
then re-measuring the blue fractions. We find that the fluctuated blue
fractions are systematically larger than the original values, since
there is a greater red population than blue around the demarcation
$U - V = 1.0$, and correct for the effect (the correction amounts are
included in the final errors). We plot the measured blue fractions as a
function of redshift and stellar mass in Fig. 10, and also list them in
Table 9. One can see that the blue fractions are significantly lower in
more massive galaxies, and that the fractions in massive systems decrease
toward the local Universe. The blue fractions in $< 10^{11.0} M_{\odot}$
galaxies increase from z = 0.4--0.6 to 0.2--0.4 because the dominant
population within the sample shifts to bluer, less massive galaxies
toward the local Universe due to the fainter detection limit. In fact
we observe a decreasing blue fraction toward the local Universe, as
seen in massive galaxies, if we take the subsample with stellar mass
$10^{10.5} M_{\odot}< M_{\star} < 10^{11.0} M_{\odot}$ (see Table 9). As discussed above,
the $10^{11.0-11.5} M_{\odot}$ galaxies at $z$ = 0.2--0.4 and the $10^{11.5-12.0} M_{\odot}$
galaxies in all redshift bins could have considerable fractions of
contamination from less massive galaxies, which likely have bluer
$U - V$ colours. Actually, investigating the spectroscopic sample
shows that the contamination makes the mean $U - V$ colours bluer
by up to 0.1 mag. Thus the true blue fractions in the above classes
of galaxies could be even smaller than the present measurements.
The lower blue fractions in more massive galaxies and the decreasing
trend toward the local Universe implies major star formation
at higher redshifts, which is in line with 'downsizing' of the star
formation \citep[e.g.,][]{cowie96}.

The above measurements suggest that the majority of massive
galaxies are fairly quiescent, while of the rest a considerable fraction
of galaxies are experiencing active star formation, especially at
higher redshifts ($\sim$30 per cent at $z \sim 1$). Such active star formation
in massive galaxies is also reported in \citet{conselice07}, who
find that nearly half of their massive ($M_{\star} > 10^{11} M_{\odot}$) galaxies at
$0.4 < z < 1.4$ are detected in the {\it Spitzer Space Telescope}/MIPS
24-$\mu$m band and the average star-formation rate amounts to
$\sim50 M_{\odot}$ yr$^{-1}$. Our measurements of blue fractions indicate that the
star-formation activity in massive galaxies is gradually quenched toward
the local Universe, leaving the most massive galaxies on the
red sequence. Star-formation quenching processes above a certain
stellar mass limit are actually proposed, such as the internal feedback
of mass assembly caused by active galactic nuclei \citep[e.g.,][]{silk98, granato04, springel05}. 
The above scenario is consistent with our primary results
that the active bottom-up formation of massive galaxies is going on
during $0 < z < 1$.

\begin{table}
 \caption{The blue population fraction.}
 \label{tab:frac_blue}
 \begin{tabular}{@{}lccc}
 \hline
             &                                 &   log $M_{\star}$ &           \\
  Redshift   & $<$ 11.0 (10.5 -- 11.0)            & 11.0 -- 11.5    & 11.5 -- 12.0\\
  \hline
  0.2 -- 0.4 & 0.54 $\pm$ 0.01 (0.36 $\pm$ 0.01) & 0.12 $\pm$ 0.05 & $<$ 0.06\\
  0.4 -- 0.6 & 0.47 $\pm$ 0.01 (0.43 $\pm$ 0.01) & 0.25 $\pm$ 0.03 & 0.18 $\pm$ 0.03\\
  0.6 -- 0.8 & 0.51 $\pm$ 0.01 (0.43 $\pm$ 0.01) & 0.28 $\pm$ 0.02 & 0.19 $\pm$ 0.06\\
  0.8 -- 1.0 & 0.59 $\pm$ 0.01 (0.56 $\pm$ 0.01) & 0.29 $\pm$ 0.02 & 0.21 $\pm$ 0.03 \\
  \hline
 \end{tabular}
\end{table}

\begin{figure}
  \includegraphics[width=84mm]{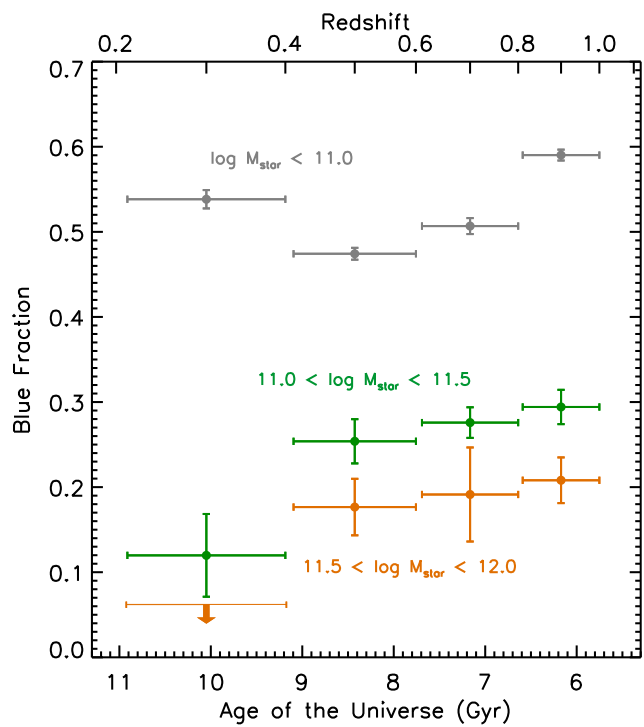}
  \caption{The fraction of the blue population in $< 10^{11.0} M_{\odot}$ (grey solid),
$10^{11.0-11.5} M_{\odot}$ (green) and $10^{11.5-12.0} M_{\odot}$ (orange) galaxies, respectively,
as a function of redshift.}
  \label{cmd2}
\end{figure}

Finally, we comment on the compatibility of the present results
with previous studies. There are a number of studies of luminous
red galaxies (LRGs) at $z < 1$ \citep[e.g.,][]{brown07, brown08, cool08} 
covering up to $\sim$10 deg$^2$. Authors consistently suggest
that the LRGs show little evolution in number density since $z \sim 1$.
However, the mass-to-optical luminosity ratio of galaxies has a significant
scatter even for the massive systems, so that galaxies with a
certain stellar mass are not quite the equivalent population of galaxies
with a certain optical luminosity. This leads to a consequence of
most significance for the most massive galaxies: a small portion of
the less massive, much more numerous galaxies could contaminate
the luminous class of galaxies if their mass-to-luminosity ratios
were slightly less than the average, and thus could easily dominate
the luminous population. Therefore a subtle (in absolute amplitude)
change in the number of most massive galaxies could be drowned
out in the measured evolution of the LRGs.

Studies also exist of massive galaxies at $z < 1$ with stellar mass
measurements based on infrared photometry \citep[e.g.,][]{conselice07, ilbert09}, 
although these studies cover a much smaller
field of view ($\la$ 1.5 deg$^2$) than ours. In contrast to the present results,
they report little evolution in number of the most massive
($> 10^{11.5} M_{\odot}$) galaxies. At least part of the discrepancy could be
due to small-number statistics and cosmic variance. Actually, while
we find $\sim$1000 samples of the most massive galaxies in each redshift
bin from our 55.2 deg$^2$, the number of samples observed over
$\sim$1.5 deg$^2$ should be only $\sim$30. We estimate the effects of cosmic
variance by dividing our total field into small subfields, each
covering $\sim$ 1.5 deg$^2$, and measure the number-density fluctuations
of massive galaxies among the subfields. As a result, we find that
the number densities of the most massive galaxies measured over
$\sim$ 1.5 deg$^2$ can fluctuate by up to a factor of a few. However, we
are not sure whether the above uncertainties alone can account
for the discrepancy between the present results and previous ones.
The number-density measurements at the steep high end of the
galaxy stellar mass function could be heavily affected by contamination
arising from less massive galaxies, thus quite accurate analysis
is required to unveil the subtle evolution of the most massive
galaxies.

In essence, our measurements provide a unique opportunity to
investigate the mean properties and evolution of the most massive
galaxies, owing to the reliable estimates of photometric redshifts
and stellar masses conducted over an unprecedentedly large field
of view. What is observationally clear is that we have discovered
a substantial deficit of the most massive galaxies out to $z = 1$
compared with the local Universe. The analysis of the rest-frame
$U - V$ colour distributions indicates that star-formation activity
might be responsible for the active build-up of these systems, while
it is possible that a so-called dry merger \citep[e.g.,][]{bell04, vandokkum05} 
is the main driver of the evolution. Actually, some
observations suggest that a substantial fraction of massive early-type
galaxies go through active evolution in terms of the galaxy structure
as well as the star formation since $z \sim$ 1 \citep[e.g.,][]{treu05, vanderwel08}. 
The present results provide crucial evidence
of hierarchical galaxy formation, the missing piece of observation
required to chart a course for future theoretical models based on
$\Lambda$CDM theory.

\section{Summary}

We present an analysis of $\sim$60 000 massive galaxies with stellar
masses $10^{11} M_{\odot} < M_{\star} < 10^{12} M_{\odot}$ in an unprecedentedly large
field of view of 55.2 deg$^2$. The galaxies are drawn from the UKIDSS
Large Area Survey K-band images on the SDSS southern equatorial
stripe. We have created deep-stacked $u$, $g$, $r$, $i$ and $z$-band images
from the SDSS Supplemental and Supernova Survey image frames,
which results in $\sim$90 per cent counterparts of the $K$-band sources in
the $g$, $r$, $i$ and $z$ bands. We also exploit the redshift surveys conducted
on the SDSS southern equatorial stripe, namely the VIMOS-VLT
Deep Survey and the DEEP2 Redshift Survey, in order to obtain
accurate photometric redshifts and associated uncertainties for the
galaxies. Stellar masses are estimated by comparing the observed
broad-band colours with stellar population synthesis models.

In each of the redshift bins $z$ = 0.2--0.4, 0.4--0.6, 0.6--0.8 and
0.8--1.0, we obtain $\sim$10 000 and $\sim$1 000 galaxies with stellar masses
$10^{11.0} M_{\odot} < M_{\star} < 10^{11.5} M_{\odot}$ and $10^{11.5} M_{\odot} < M_{\star} < 10^{12.0} M_{\odot}$,
respectively. The galaxies are almost completely detected out to
$z = 1$, and form by far the largest sample of massive galaxies
reaching to the Universe at about half its present age. We find that
the most massive ($10^{11.5} M_{\odot} < M_{\star} < 10^{12.0} M_{\odot}$) galaxies have
experienced rapid growth in number since $z = 1$, while the number
densities of less massive systems show rather mild evolution.
Such a hierarchical trend of evolution is consistent with the predictions
of the current semi-analytic galaxy formation model based on
$\Lambda$CDM theory. While the majority of the massive galaxies are red-sequence
populations, we find that a considerable fraction are blue
star-forming galaxies. The blue fraction is less in more massive systems
and decreases toward the local Universe, leaving the red, most
massive galaxies at low redshifts, which further supports the idea of
active bottom-up formation of these populations during $0 < z <1$.
The present results provide strong evidence that galaxy formation
proceeds in a hierarchical way, and place stringent observational
constraints on future theoretical models.

\section*{Acknowledgments}

We are grateful to K. Shimasaku, K. Kohno, J. Makino, N. Yasuda
and N.Yoshida for insightful discussions and suggestions. We thank
the referee for many useful comments that have helped to improve
this paper. YM acknowledges Grant-in-Aid from the Research Fellowships
of the Japan Society for the Promotion of Science (JSPS)
for Young Scientists. This work was supported by Grants-in-Aid
for Scientific Research (17104002, 21840027), Specially Promoted
Research (20001003) and the Global COE Program of Nagoya University
'Quest for Fundamental Principles in the Universe (QFPU)'
from JSPS and MEXT of Japan.

This publication makes use of data products from the Two-
Micron All-Sky Survey, which is a joint project of the University
of Massachusetts and the Infrared Processing and Analysis
Center/California Institute of Technology, funded by the National
Aeronautics and Space Administration and the National Science
Foundation. IRAF is distributed by the National Optical Astronomy
Observatories, which are operated by the Association of Universities
for Research in Astronomy, Inc., under cooperative agreement
with the National Science Foundation. Funding for the SDSS and
SDSS-II has been provided by the Alfred P. Sloan Foundation, the
Participating Institutions, the National Science Foundation, the US
Department of Energy, the National Aeronautics and Space Administration,
the Japanese Monbukagakusho, the Max Planck Society
and the Higher Education Funding Council for England. The SDSS
Web Site is http://www.sdss.org/. The SDSS is managed by the Astrophysical
Research Consortium for the Participating Institutions.
The Participating Institutions are the American Museum of Natural
History, Astrophysical Institute Potsdam, University of Basel, University
of Cambridge, CaseWestern Reserve University, University
of Chicago, Drexel University, Fermilab, the Institute for Advanced
Study, the Japan Participation Group, Johns Hopkins University, the
Joint Institute for Nuclear Astrophysics, the Kavli Institute for Particle
Astrophysics and Cosmology, the Korean Scientist Group, the
Chinese Academy of Sciences (LAMOST), Los Alamos National
Laboratory, the Max-Planck-Institute for Astronomy (MPIA), the
Max-Planck-Institute for Astrophysics (MPA), New Mexico State
University, Ohio State University, University of Pittsburgh, University
of Portsmouth, Princeton University, the United States Naval
Observatory and the University of Washington.

\label{lastpage}

\end{document}